\documentclass[5p]{elsarticle}				 
\usepackage{graphicx}																												 

\begin{document}																													 
\title{Asymptotics of quasi-classical localized states in 2D system of charged hard-core bosons}

\author{Yu.D.~Panov\corref{cor1}}
\ead{yuri.panov@urfu.ru}
\author{A.S.~Moskvin}

\cortext[cor1]{Corresponding author}

\address[urfu]{Institute of Natural Sciences and Mathematics, Ural Federal University, 620002, Ekaterinburg, Russia}

\begin{abstract}
The continuous quasi-classical two-sublattice approximation is constructed for the 2D system of charged hard-core bosons to explore metastable inhomogeneous states analogous to inhomogeneous localized excitations in magnetic systems. 
The types of localized excitations are determined by asymptotic analysis and compared with numerical results. 
Depending on the homogeneous ground state,
the excitations are the ferro and antiferro type vortices, the skyrmion-like topological excitations or linear domain walls.
\end{abstract}



\maketitle
%

\section{Introduction}
\label{intro}

Unusual phase states with non-zero diagonal and off-diagonal order parameters exist in various models of lattice bosons.
Last years the interest to these models was boosted by finding of a coexistence of charge ordering and superconductivity in high temperature superconductors.
The charge degree of freedom in high-T$_c$ cuprates such as La$_{2-x}$Sr$_x$CuO$_4$ can be described in terms of a pseudospin $S=1$ model\,\cite{Moskvin2011,Moskvin2015}. 
Here in the paper we consider the limiting case of the model, reduced to the 2D system of charged hard-core (\emph{hc})  bosons on a square lattice.
Quasi-classical continuous description of the 2D magnetic systems reveals their striking features, namely, the collective localized inhomogeneous states with nontrivial topology and finite excitation energy.
These include topological solitons\,\cite{Belavin1975,Voronov1983}, magnon drops\,\cite{Ivanov1977}, in- and out-of-plane vortex-antivortex pairs\,\cite{Gouva1989}, and various spiral solutions\,\cite{Borisov2001,Bostrem2002,Borisov2004}.
Basically these solutions have been obtained for the isotropic and anisotropic ferromagnet. 
We construct a continuous two-sublattice approximation for the 2D system of charged \emph{hc} bosons, that is equivalent to the highly anisotropic $s=1/2$ 2D magnets with a constant magnetization. 
The well-known ground-states phase diagram of \emph{hc} bosons contains superfluid (SF), supersolid (SS) and charge-ordered (CO) phases. 
We consider analytically the asymptotic behavior of localized excitations which converge to homogeneous solutions at infinity. 
In the system of lattice bosons, collective localized inhomogeneous states correspond to an inhomogeneous distribution of the charge and the superfluid densities.
Asymptotic analysis shows that in the SF phase the excitations are vortices with a charge inhomogeneity of “ferro” and “antiferro” ordering type in the vortex core. 
Near the border with the SS phase, the “antiferro” type vortices begin to dominate; their inflation is preceded by a change in the homogeneous ground state from the SF to SS phase. 
In the SS phase, we find that asymptotic behavior of localized excitations is consistent with skyrmion-like solutions. 
They include coherent excitations both of the superfluid component and the boson density and result in appearance of domains of CO and SF phases. 
In the CO phase, the picture of the asymptotic behavior of localized excitation is qualitatively different from that of the SF and SS phases, being consistent with the results of numerical calculations.
The results are compared with numerical calculations of the ground-state energy in the quasi-classical approximation.


\section{The model}
The $S=1$ pseudospin formalism\,\cite{Moskvin2011,Moskvin2015} developed for the  copper oxides implies that the on-site Hilbert space  described by three effective valence states of the copper ions, Cu$^{1+,2+,3+}$, or, strictly speaking, of the copper-oxygen clusters CuO$_4^{7-,6-,5-}$ in the CuO$_2$ plane. 
These states correspond to the hole occupation numbers $n_h = 0,\, 1,\, 2$  
and can be considered as components of the pseudospin $S=1$ triplet with projections $M_S = {-}1,\, 0,\, {+}1$, respectively.
A simplified pseudospin Hamiltonian that takes into account only the two-hole transport is
$$
		\mathcal{H} 
		= \Delta \sum_i S_{iz}^2
		+ V \sum_{\left\langle ij\right\rangle} S_{iz} S_{jz} 
		- t \sum_{\left\langle ij\right\rangle} \big( S_{i+}^2 S_{j-}^2 + S_{i-}^2 S_{j+}^2 \big)
		.
$$
A pair of holes plays the role of one boson in the model. 
Here we consider the limiting case $\Delta\rightarrow-\infty$, or the large negative $U$ limit.
This condition eliminates the on-site state with $n_h = 1$, 
so we come to the Hamiltonian of charged \emph{hc} bosons in terms of a pseudospin $\vec{\sigma}$, $\sigma_z = \pm 1$:
\begin{equation}
		\mathcal{H} =  
		- t \sum_{\left\langle ij\right\rangle} \big( \sigma_{i+} \sigma_{j-} + \sigma_{i-} \sigma_{j+} \big)
		+ V \sum_{\left\langle ij\right\rangle} \sigma_{iz} \sigma_{jz} 
	.
\end{equation}
Here $\sigma_\alpha$, $\alpha=x,y,z$ are Pauli matrices, $\sigma_{\pm}=(\sigma_x \pm i\sigma_y)/2$.
The $z$-component of the pseudospin describes the local density of bosons, 
so that antiferromagnetic $z$-$z$ exchange corresponds to the repulsive density-density interaction, 
while isotropic ferromagnetic planar exchange corresponds to the kinetic energy of bosons. 
The constant total number of bosons leads to the constraint on the total $z$-component of the pseudospin.
We define the $n$, as the density of the total doped charge counted from the state with a zero total z-component, 
or parent Cu$^{2+}$ state.
Then $n$ is the sum of $z$-components of the pseudospin:  $\sum_i \sigma_{iz} = nN$, 
where $N$ is the total number of sites.
If  $\rho$ is the density of \emph{hc} bosons, then  $n$ is the deviation from the half-filling:
$\rho = (1+n)/2$.

The energy functional 
$E = \left\langle \Psi \left| \mathcal{ H }\right| \Psi \right\rangle$ 
in a quasi-classical approximation with
\begin{equation}
	\left| \Psi \right\rangle
	= \prod_i
	\bigg( 
		\cos \frac{\theta_i}{2} \, e^{-i\frac{\phi_i}{2}} \, | {+}1 \rangle
		+ \sin \frac{\theta_i}{2} \, e^{ i\frac{\phi_i}{2}} \, | {-}1 \rangle
	\bigg)
	,
	\label{EnF}
\end{equation}
where $| \pm1 \rangle$ are the eigenfunctions of the $\sigma_{z}$ on $i$-th site, takes the form
\begin{eqnarray}
	\varepsilon 
	&=& 
	- \sum_{\left\langle i,j\right\rangle} \sin\theta_i \, \sin\theta_j \, \cos(\phi_i-\phi_j)
	+{}
	\label{en}
	\\
	&&
	{}+ \lambda \sum_{\left\langle i,j\right\rangle} \cos\theta_i \, \cos\theta_j
	- \xi  \bigg( \sum_i \cos\theta_i   -   n N \bigg)
	.
	\nonumber
\end{eqnarray}
Here $\left\langle i,j\right\rangle$ denotes summation over nearest neighbors in a square lattice. 
The $\theta_i$ and  $\phi_i$ are the polar and azimuthal angles of the quasiclassical pseudospin vector at an $i$-th site.
We define
$\varepsilon = 2{E}/t$,
$\lambda = 2V/t$,
$\xi = 2\mu/t$,
where the chemical potential $\mu$ takes into account the bosons density constraint.

Hereafter, we introduce two sublattices $A$ and $B$ with the checkerboard ordering. 
The first sum in the Exp.(\ref{en}) has its lowest value if ${\cos(\phi_i-\phi_j)=1}$.
This allows us to make a simplifying assumption, that $\phi_A(\vec{r}) = \phi_B(\vec{r}) \equiv \phi(\vec{r})$. 
It is worth to note that this assumption is confirmed by the results of our numerical simulations.
We define functions 
$	u(\vec{r}) \equiv \theta_A(\vec{r}) $, 
$	v(\vec{r}) \equiv \theta_B(\vec{r}) $, 
and their combinations 
$f = \cos u \cos v$, $F = - f + \lambda f_{uv}$, 
where subscripts $u$ and $v$ denote derivatives with respect to these quantities.
Then the Euler equations in the continuous approximation take a compact form
\begin{equation}
	\!\!
	\!\!
	\!\!
	\!\!
	\!\!
	\!\!
	\left\{
	\!\!
	\begin{array}{l}
		f_{uv} \Delta \phi  -  2 f_v \cdot ( \nabla u , \nabla \phi )  =  0
		,\\
		f_{uv} \Delta \phi  -  2 f_u \cdot ( \nabla v , \nabla \phi )  =  0
		,\\
		\displaystyle
		F \Delta u  +  F_u \cdot ( \nabla u )^2  -  f_u \cdot ( \nabla \phi )^2  -  4F_u  +  \xi \, \sin v  =  0
		,\\
		\displaystyle
		F \Delta v  +  F_v \cdot ( \nabla v )^2  -  f_v \cdot ( \nabla \phi )^2  -  4F_v  +  \xi \, \sin u  =  0
		.
	\end{array}
	\right.
	\!\!
	\label{EqEulerSys1}
\end{equation}
These equations need to add the bosons density constraint. 
With the relevant exchange constants, the equations (\ref{EqEulerSys1}) lead to the equations of \cite{Egorov2002}.


\section{The asymptotic behavior of localized solutions}
\label{sec-as}

The system (\ref{EqEulerSys1}) along with the boson density constraint has constant (homogeneous) solutions, 
$\phi=\phi_0$, $u=u_0$, $v=v_0$, 
that determine the well-known ground-state phase diagram\,\cite{Micnas1990}  
of the \emph{hc} bosons system in the mean-field approximation.

Given $\lambda<1$ or $n^2>(\lambda-1)/(\lambda+1)$ the ground state of the system is a superfluid (SF)
with
	$\cos u_0 = \cos v_0 = n$, 
	$\varepsilon_0 = -2 + 2 \left(\lambda + 1\right) n^2$.
Given $n^2<(\lambda-1)/(\lambda+1)$ the ground state is a supersolid (SS) with
	$\cos u_0 = n + z$,
	$\cos v_0 = n - z$,
where $z^2 = 1 + n^2 - 2|n|\lambda/\sqrt{\lambda^2 - 1}$, and
	$\varepsilon_0 = -2\lambda + 4|n|\sqrt{\lambda^2 - 1}$.
In all phases, the value of $\xi$ satisfies the regular expression $\xi_0 = \partial \varepsilon_0 / \partial n$.
When $\lambda>1$ and $n = 0$ the SS phase transforms into a conventional charge-ordered (CO) phase with the checkerboard ordering.

\begin{figure}
	\centering
		\includegraphics[width=0.48\textwidth]{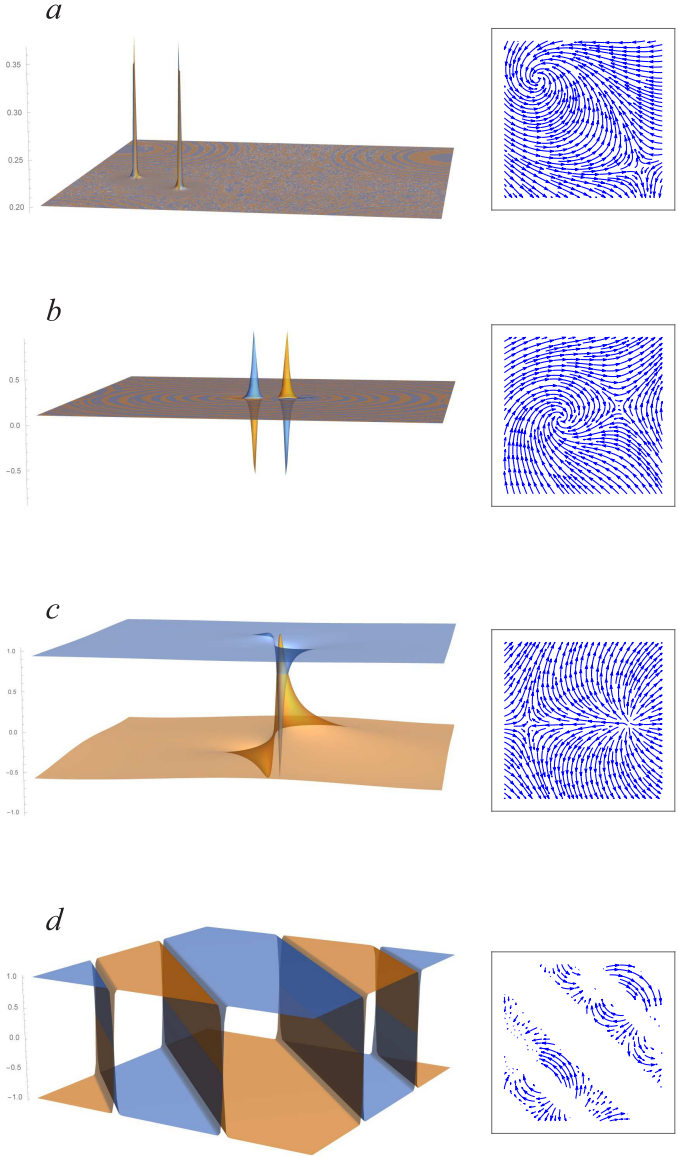}
	\caption{(color online)
	Inhomogeneous states in 2D system of charged hard-core bosons. 
	The left panels show 
	local charge density $n_i = \sigma_{zi} = \cos \theta_i$.
	The difference of sublattice states is clearly evident in the cases b, c and d.
	The right panels show 
	phase flow of planar components of the pseudospin, $\sigma_{xi}$ and $\sigma_{yi}$.
	The phase flow reveals the vortex-antivortex pair structure
	in the core of inhomogeneity of the local charge density in the cases a, b and c.
	The parameters of the model are: 
	a) $n=0.2$, $\lambda=0.5$ (SF phase);
	b) $n=0.1$, $\lambda=0.9$ (SF phase);
	c) $n=0.2$, $\lambda=1.5$ (SS phase);
	d) $n=0.0$, $\lambda=1.5$ (CO phase). 
	These sets of parameters are denoted with letters \emph{a}-\emph{d}  on the ground-state phase diagram in Fig.\ref{fig:AsPhaseDiag}.
	} 
	\label{fig:1}
\end{figure}

The equations (\ref{EqEulerSys1}) in the case of $\lambda <0$ have localized solutions with nonzero topological charge and finite energy \cite{Belavin1975,Voronov1983,Ivanov1977,Gouva1989,Borisov2001,Bostrem2002,Borisov2004}.
However, in our case $\lambda>0$, numerical calculations with the conjugate gradient method for 
minimizing of the energy functional (\ref{EnF}) 
on the lattice 256$\times$256 indicate the existence of similar solutions, at least, as metastable states.
The results are shown in Fig.\ref{fig:1}.
The actual stability of these solutions in our calculation was different. 
The SF-phase solutions, similar to that of in Fig.\ref{fig:1}, cases \emph{a} and \emph{b}, quickly evolved to an uniform one. 
The SS- and CO-phase solutions, similar to that of in Fig.\ref{fig:1}, cases \emph{c} and \emph{d}, 
retain its form for more than $10^6$ iterations.

We investigated the asymptotic behavior of localized solutions, suggesting that at $r\rightarrow\infty$ they have the form
$\phi(\vec{r}) = \phi_0 + \tilde{\phi}(\vec{r})$,
$u(\vec{r}) = u_0 + \tilde{u}(\vec{r})$, 
$v(\vec{r}) = v_0 + \tilde{v}(\vec{r})$, 
where $\tilde{\phi},\tilde{u},\tilde{v}\rightarrow0$ at $r\rightarrow\infty$. 
Hereinafter, the index 0 means the corresponding values for constant solutions. 
The linearized system (\ref{EqEulerSys1}) for the functions $\tilde{\phi},\tilde{u},\tilde{v}$ takes the form
\begin{equation}
	\left\{
	\begin{array}{l}
		\displaystyle
		f_{uv0} \Delta \tilde{\phi}  =  0
		,\\
		\displaystyle
		F_0 \Delta \tilde{u} 
		+ 4  F_0  \tilde{u} 
		+ 4  \left( - F_{uv0}  +  \frac{\xi_0}{4} \cos v_0 \right)  \tilde{v}  
		=  0
		,\\
		\displaystyle
		F_0 \Delta \tilde{v}  
		+ 4  F_0  \tilde{v} 
		+ 4  \left( - F_{uv0}  +  \frac{\xi_0}{4} \cos u_0 \right)  \tilde{u}
		=  0
		.
	\end{array}
	\right.
	\label{EqEulerSys2}
\end{equation}

In the case of the SF and SS phases, the solutions for the first equation can be written as
\begin{equation}
	\tilde{\phi}(\vec{r}) 
	= \sum_{m=1}^{\infty}  \frac{c_m}{r^m}  \cos m(\varphi-\varphi_m)
	,
	\label{phimult}
\end{equation}
with $c_m$ and $\varphi_m$ determined by the boundary conditions. 
In the case of the CO phase, the first equation reduces to an identity since $f_{uv0}=0$.

In the case of the SF phase, the second and the third equations become independent Helmholtz equations 
for the ferro- and antiferro-type combinations 
$U = \tilde{u} + \tilde{v}$, $V = \tilde{u} - \tilde{v}$:
\begin{equation}
		\Delta U + A_1 \, U  =  0
		,\quad
		A_1 = 4 \frac{(\lambda+1)(1-n^2)}{\lambda-(\lambda+1)n^2}
		;
		\label{linEqsSFU}
\end{equation}
\begin{equation}
		\Delta V + A_2 \, V  =  0
		,\quad
		A_2 = 4 \frac{\lambda-1-(\lambda+1)n^2}{\lambda-(\lambda+1)n^2}
		.
		\label{linEqsSFV}
\end{equation}

The solutions has the form
\begin{eqnarray}
	\Phi(\vec{r},r_i)
	&=&
	\sum_{l=0}^{\infty}   a_{l}  K_l ( r/r_i )  \cos l ( \varphi - \alpha_{l} )
	,
	\label{Phi}
	\\
	\Psi(\vec{r},r_i)
	&=&
	\sum_{l=0}^{\infty} 
	\Big[ 
		b_{1l}  J_l ( r/r_i ) \cos l (\varphi-\beta_{1l})
		+{}
		\nonumber
	\\
	&&	
	{}+ b_{2l}  Y_l ( r/r_i ) \cos l (\varphi-\beta_{2l})
	\Big]
	,	
	\label{Psi}
\end{eqnarray}
where $K_l$ are the Macdonald functions, 
$J_l$ and $Y_l$ are the Bessel functions of the first and second kind, 
and $a_{l}$, $\alpha_{l}$, $b_{kl}$, $\beta_{kl}$, $k=1,2$ are some constants.
An analysis of the asymptotic behavior of solutions (\ref{Phi},\ref{Psi}) and the requirement that the omitted nonlinear terms in equations (\ref{EqEulerSys2}) are small in comparison with the remaining linear terms determine that $\Psi = 0$.
The account in the lowest order of the mixing with the function $\tilde{\phi}$ 
does not change $V(\vec{r})$ and gives additional term in $U(\vec{r})$
having asymptotic behavior:
\begin{equation}
	U_1(\vec{r}) \approx - \frac{n c_m^2}{2(\lambda+1)\sqrt{1-n^2}}  \frac{m^2}{r^{2m+2}}
	,
	\label{U1SF}
\end{equation}
where $m$ is the number that specifies first nonzero term in (\ref{phimult}).

Line $n^2 = \lambda/(\lambda+1)$ is the boundary of areas of the SF-phase with a different behavior of the  $U$ and $V$ functions
\begin{equation}
		n^2 > \frac{\lambda}{\lambda+1}: 
		\;
		U(\vec{r}) = \Phi(\vec{r},r_1) + U_1(\vec{r})
		,\;
		V(\vec{r}) = 0
		;
		\label{UVfin1}
\end{equation}
\begin{equation}
		n^2 < \frac{\lambda}{\lambda+1}:
		\;
		U(\vec{r}) = U_1(\vec{r})
		,\;\;
		V(\vec{r}) = \Phi(\vec{r},r_2)
		.
		\quad\;
		\label{UVfin2}
\end{equation}
Here we define characteristic lengths $r_i^{-2} = |A_i|$.

In the case of the SS phase, we need to define the ferri-type combinations: 
$\tilde{U} = A \tilde{u} + \tilde{v}$ and $\tilde{V} = A \tilde{u} - \tilde{v}$, 
where $A = -F_{uv0} + \frac{\xi_0}{4} \cos u_0$.
The equations (\ref{EqEulerSys2}) lead to Helmholtz equation for the $\tilde{U}$ function having solution
$\Psi(\vec{r},r_3)$,
$r_3^{-2} = 8$. 
As in previous case we have to put $\Psi = 0$.
The $\tilde{V}$ function obeys to Laplace equation. 
Taking into account the mixing  in the lowest order  with the function $\tilde{\phi}$ we come to the expressions
\begin{equation}
		\tilde{U}(\vec{r})  
		= \frac{ c_m^2 }{8} \left( A f_{u0} + f_{v0} \right) \frac{m^2}{r^{2m+2}}
		,
\end{equation}
\begin{eqnarray}
		\tilde{V}(\vec{r}) 
		&=&
		\sum_{l=1}^{\infty}  \frac{C_l}{r^l}  \cos l (\varphi-\gamma_l)
		+{}
		\nonumber
		\\
		&&
		{}+
		\frac{ c_m^2 }{4} \left( A f_{u0} - f_{v0} \right) \frac{1}{r^{2m}}
		,
		\label{VSS}
\end{eqnarray}
where $m$ is the number that specifies first non-zero term in (\ref{phimult}), and the expressions 
$f_{u0}=-\sin u_0 \cos v_0$, $f_{v0}=-\cos u_0 \sin v_0$ are determined by the expressions for the constant solutions.

Similarly the case of the CO phase, 
we obtain
\begin{equation}
	\tilde{\phi}(\vec{r}) = 0
	,\quad
	U(\vec{r})= 0
	,\quad
	V(\vec{r}) = \Phi(\vec{r},r_4)
	,
	\label{COUV}
\end{equation}
where 	$r_4^{-2} = 4  (  \lambda - 1  )$.


\section{Discussion}
\label{sec-diss}

\begin{figure}
	\centering
		\includegraphics[width=0.5\textwidth]{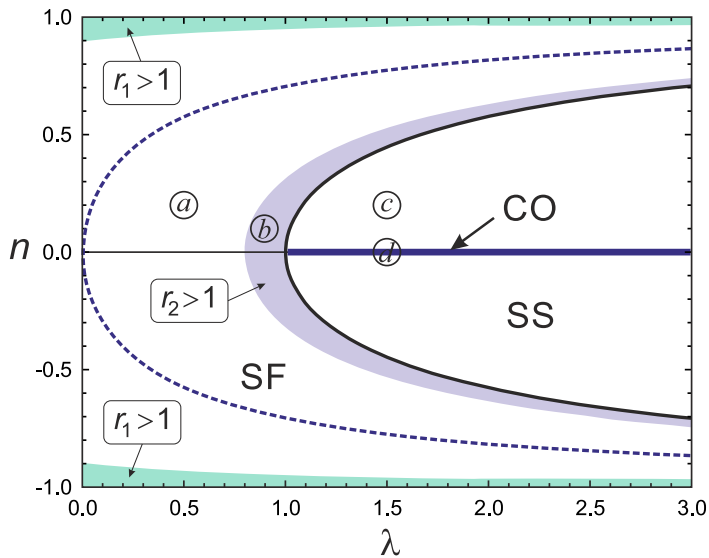}
		\caption{(color online)
		The ground state phase diagram of the \emph{hc} bosons in the mean-field approximation.
		The solid line corresponds to SF-SS phase boundary, $n^2 = (\lambda-1)/(\lambda+1)$. 
		The thick line at $\lambda>1$, $n=0$ shows CO phase.
		The dotted line, $n^2 = \lambda/(\lambda+1)$, 
		separates the two types of the asymptotic behavior in accordance with the expressions (\ref{UVfin1}) and (\ref{UVfin2}).
		In the shaded areas inside the SF phase region 
		the characteristic lengths satisfy to inequalities $r_i>1$. 
		The letters \emph{a}-\emph{d} in the circles correspond to the parameters sets in Fig.\ref{fig:1}.
		}
	\label{fig:AsPhaseDiag}
\end{figure}

The analysis of the asymptotic behavior of the localized states reveals qualitative differences of the finite energy excitations in the SF, SS, and CO phases. 

In the SF phase an asymptotic of the polar angle of the pseudospin vector  is determined by the expressions (\ref{UVfin1}, \ref{UVfin2}).
When comparing these results with numerical calculations it is worth to note that the characteristic lengths obey to inequality $r_i<1$ in the most part of the phase diagram in the SF phase except for the areas indicated shadowed in Fig.\ref{fig:AsPhaseDiag}, so the function $\Phi$ goes to zero value  
very fast 
with increasing of $r$. 
On the contrary, the asymptotic behavior of the azimuthal angle of the pseudospin (\ref{phimult}) has no characteristic scale. 
This means that in the SF phase the main excitations are almost in-plane vortex-antivortex pairs. 
They have well localized out-of-plane core of the ferro type, with $\sigma_{zA}=\sigma_{zB}$, as shown in Fig.\ref{fig:1}a,
and become the pure in-plane ones at $n=0$ in accordance with expression (\ref{U1SF}).
The same type of localized solutions was found by the authors of \cite{Gouva1989}.
For the \emph{hc} bosons, the polar angle is related with the density of bosons, while the azimuthal angle is responsible for the superfluid density, hence these states correspond to the excitation of the superfluid component with highly localized  heterogeneity of bosons density in the foci of the vortex-antivortex pairs.
In the shaded region in the SF phase near the border of the SF-SS phases in Fig.\ref{fig:AsPhaseDiag}, the antiferro type vortices (see Fig.\ref{fig:1}b), with $\sigma_{zA}\neq\sigma_{zB}$, begin to dominate, their inflation  is preceded by a change of the homogeneous ground state from SF to SS phase. 

The CO phase has no linear excitation of $\tilde{\phi}$. The characteristic lengths of the azimuthal excitations (\ref{COUV}) are small except the region near $\lambda=1$. 
This results in a high stability of the homogeneous CO phase. 
A typical picture of the inhomogeneous state shown in Fig.\ref{fig:1}d is represented by linear domains of the CO phase. 
The non-zero values of the SF order parameter are realized within the domain walls, thus  
giving rise to appearance of a filamentary superfluidity in the \emph{hc} bosons.

In the SS phase the asymptotic behavior of the polar (\ref{phimult}) and azimuthal (\ref{VSS}) excitations is qualitatively the same without characteristic scales. 
Hence in this case there are skyrmion-like excitations as shown in Fig.\ref{fig:1}c. 
For the \emph{hc} bosons these coherent states include both the excitations of the superfluid component and the boson density. 
In a center of skyrmion the difference $\sigma_{zA}-\sigma_{zB}$ has maximal magnitude, that corresponds to CO phase, 
and near there is a region where $\sigma_{zA}-\sigma_{zB}=0$, that corresponds to SF phase. 
So, the skyrmion-like excitations in the SS phase generate the topological phase separation. 
Note, that another type of instability in SS phase were also found by Quantum Monte Carlo calculations\,\cite{Batrouni2000}.

\quad

Funding: The work supported by Act 211 Government of the Russian Federation, agreement No 02.A03.21.0006 and by the Ministry of Education and Science, projects 2277 and 5719.


\end{document}